\documentclass[pre,twocolumn,showpacs,superscriptaddress,preprintnumbers,floatfix]{revtex4-1}

\usepackage{etex}
\usepackage{ifpdf}
\usepackage{hyperref}
\usepackage{dcolumn}
\usepackage{url}
\usepackage{amsmath}
\usepackage{amscd}
\usepackage{amsfonts}
\usepackage{amssymb}
\usepackage{bm}   
\usepackage{bbm}
\usepackage{verbatim}
\usepackage{stmaryrd}
\usepackage{amsthm}
\usepackage{xcolor}
\usepackage{setspace}
\usepackage{subfigure}

\ifpdf
  \usepackage[pdftex]{graphicx}
\else
  \usepackage[dvips]{graphicx}
\fi

\newif \ifgenerate

\generatefalse

\ifgenerate

  \usepackage{tikz}
  \usetikzlibrary{external}
  \usetikzlibrary{arrows}
  \usetikzlibrary{automata}
  \usetikzlibrary{positioning}

  \tikzstyle{vaucanson}=[
    node distance=2.5cm,
    bend angle=15,
    auto,
    every loop/.style={},
    every edge/.style={->,draw=black,line width=1.2,>=latex,shorten <=1pt,shorten >=1pt},
    every state/.style={draw=black,line width=2,font=\large},
    loop right/.style={right,out=22,in=-22,loop},
    loop above/.style={above,out=112,in=68,loop},
    loop left/.style={left,out=202,in=158,loop},
    loop below/.style={below,out=292,in=248,loop},
    loop above right/.style={right,out=67,in=23,loop},
    loop above left/.style={left,out=157,in=113,loop},
    loop below left/.style={left,out=247,in=203,loop},
    loop below right/.style={right,out=337,in=293,loop},
  ]

\fi

\newcommand{\Symbol}[1]{\textcolor{blue}{#1}}
\newcommand{\Edge}[2]{#1|\Symbol{#2}}

\theoremstyle{plain}	
\theoremstyle{plain}	
\theoremstyle{plain} 	
\theoremstyle{plain} 	
\theoremstyle{plain} 	
\theoremstyle{plain} 	
\theoremstyle{plain} 	\newtheorem{Prop}{Proposition}
\theoremstyle{plain} 	
\theoremstyle{plain} 	
\theoremstyle{plain}	
\theoremstyle{plain}	
\theoremstyle{plain}	

\def\clap#1{\hbox to 0pt{\hss#1\hss}}

\def\mathclap{\mathpalette\mathclapinternal}

\def\mathclapinternal#1#2{%
\clap{$\mathsurround=0pt#1{#2}$}}

\newcommand{\eM}     {\mbox{$\epsilon$-machine}}
\newcommand{\eMs}    {\mbox{$\epsilon$-machines}}
\newcommand{\EM}     {\mbox{$\epsilon$-Machine}}

\newcommand{\MeasAlphabet}  {\mathcal{A}}
\newcommand{\MeasSymbol}   { {X} }

\newcommand{\MeasSymbols}[2]{ \MeasSymbol_{#1:#2} }

\newcommand{\Past} { \MeasSymbols{}{0} }
\newcommand{\Future} { \MeasSymbols{0}{} }

\newcommand{\CausalState}   { \mathcal{S} }

\newcommand{\Prob}      {\Pr} 

\newcommand{\hmu}       {h_\mu}
\newcommand{\EE}        {{\bf E}}

\newcommand{\forward}{+}
\newcommand{\reverse}{-}
\newcommand{\forwardreverse}{\pm} 

\newcommand{\FutureCausalState} { {\CausalState}^{\forward} }

\newcommand{\PastCausalState}   { {\CausalState}^{\reverse} }

\newcommand{\lastindex}[2]{
  \edef\tempa{0}
  \edef\tempb{#2}
  \ifx\tempa\tempb
    \edef\tempc{#1}
  \else
    \edef\tempa{0}
    \edef\tempb{#1}
    \ifx\tempa\tempb
      \edef\tempc{#2}
    \else
      \edef\tempc{#1+#2}
    \fi
  \fi
  \tempc
}

\newcommand{\rhomu}{\rho_\mu}
\newcommand{\rmu}{r_\mu}
\newcommand{\bmu}{b_\mu}
\newcommand{\qmu}{q_\mu}
\newcommand{\wmu}{w_\mu}
\newcommand{\imu}{i_\mu}
\newcommand{\sigmamu}{\sigma_\mu}
\newcommand{\EER}{\EE_R}
\newcommand{\EEB}{\EE_B}
\newcommand{\EEQ}{\EE_Q}
\newcommand{\EEW}{\EE_W}
\newcommand{\I}{\mathbf{I}}

\newcommand{\CSjoint}[1][,]{
   \edef\tempa{:}
   \edef\tempb{#1}
   \ifx\tempa\tempb
      \ensuremath{\FutureCausalState\!#1\PastCausalState}
   \else
      \ensuremath{\FutureCausalState#1\PastCausalState}
   \fi
}

\newif\ifpm
\edef\tempa{\forwardreverse}
\edef\tempb{\pm}
\ifx\tempa\tempb
   \pmfalse
\else
   \pmtrue
\fi

\begin{document}

\title{Anatomy of a Bit:\\
Information in a Time Series Observation}

\author{Ryan G. James}
\email{rgjames@ucdavis.edu}
\affiliation{Complexity Sciences Center}
\affiliation{Physics Department\\
University of California at Davis,\\
One Shields Avenue, Davis, CA 95616}

\author{Christopher J. Ellison}
\email{cellison@cse.ucdavis.edu}
\affiliation{Complexity Sciences Center}
\affiliation{Physics Department\\
University of California at Davis,\\
One Shields Avenue, Davis, CA 95616}

\author{James P. Crutchfield}
\email{chaos@cse.ucdavis.edu}
\affiliation{Complexity Sciences Center}
\affiliation{Physics Department\\
University of California at Davis,\\
One Shields Avenue, Davis, CA 95616}
\affiliation{Santa Fe Institute\\
1399 Hyde Park Road, Santa Fe, NM 87501}

\date{\today}

\bibliographystyle{unsrt}

\begin{abstract}
  Appealing to several multivariate information measures---some familiar, some
  new here---we analyze the information embedded in discrete-valued stochastic
  time series. We dissect the uncertainty of a single observation to demonstrate
  how the measures' asymptotic behavior sheds structural and semantic light on
  the generating process's internal information dynamics. The measures scale
  with the length of time window, which captures both intensive (rates of
  growth) and subextensive components. We provide interpretations for the
  components, developing explicit relationships between them. We also identify
  the informational component shared between the past and the future that is not
  contained in a single observation. The existence of this component directly
  motivates the notion of a process's effective (internal) states and indicates
  why one must build models.

\vspace{0.1in}


\end{abstract}

\preprint{Santa Fe Institute Working Paper 11-05-XXX}
\preprint{arxiv.org:1105.XXXX [physics.gen-ph]}

\maketitle
\

\setstretch{1.1}

{\bf
A single measurement, when considered in the context of the past and the
future, contains a wealth of information, including distinct kinds of
information. Can the present measurement be predicted from the past? From the
future? Or, only from them together? Or not at all? Is some of the measurement
due to randomness? Does that randomness have consequences for the future or it
is simply lost? We answer all of these questions and more, giving a complete
dissection of a measured bit of information.
}

\section{Introduction}
\label{sec:introduction}

In a time series of observations, what can we learn from just a single
observation? If the series is a sequence of coin flips, a single observation
tells us nothing of the past nor of the future. It gives a single bit of
information about the present---one bit out of the infinite amount the time
series contains. However, if the time series is periodic---say, alternating $0$s
and $1$s---then with a single measurement in hand, the entire observation series need not be stored; it can be
substantially compressed. In fact, a single observation tells us the
oscillation's phase. And, with this single bit of information, we have learned
\emph{everything}---the full bit that the time series contains. Most systems
fall somewhere between these two extremes. Here, we develop an analysis of the
information contained in a single measurement that applies
across this spectrum.

Starting from the most basic considerations, we deconstruct what a measurement
is, using this to directly step through and preview the main results. With that
framing laid out, we reset, introducing and reviewing the relevant tools
available from multivariate information theory including several that have been
recently proposed. At that point, we give a synthesis employing information
measures and the graphical equivalent of the information diagram. The result is
a systematic delineation of the kinds of information that the distribution of
single measurements can contain and their required contexts of interpretation.
We conclude by indicating what is missing in previous answers to the measurement
question above, identifying what they do and do not contribute, and why
alternative state-centric analyses are ultimately more comprehensive.

\section{A Measurement: A Synopsis}
\label{sec:anatomy-bit}

For our purposes an instrument is simply an interface between an observer and
the system to which it attends. All the observer sees is the instrument's
output---here, we take this to be one of $k$ discrete values. And, from a series
of these outputs, the observer's goal is to infer and to understand as much
about the system as possible---how predictable it is, what are the active
degrees of freedom, what resources are implicated in generating its behavior,
and the like.

The first step in reaching the goal is that the observer must store at least one
measurement. How many decimal digits must its storage device have? To specify
which one of $k$ instrument outputs occurred the device must use $\log_{10} k$
decimal digits. If the device stores binary values, then it must provide $\log_2
k$ bits of storage. This is the maximum for a one-time measurement. If we
perform a series of $n$ measurements, then the observer's storage device must
have a capacity of $n \log_2 k$ bits.

Imagine, however, that over this series of measurements it happens that output
$1$ occurs $n_1$ times, $2$ occurs $n_2$ times, and so on, with $k$ occurring
$n_k$ times. It turns out that the storage device can have much less capacity;
using less, sometimes substantially less, than $n \log_2 k$ bits.

To see this, recall that the number $M$ of possible sequences of $n$
measurements with $n_1, n_2, \ldots, n_k$ counts is given by the multinomial
coefficient:
\begin{align*}
M & = {n \choose{n_1~n_2~\cdots~n_k}} \\
  & = \frac{n!}{n_1 ! \cdots n_k !} ~.
\end{align*}
So, to specify which sequence occurred we need no more than:
\begin{align*}
 k \log_2 n + \log_2 M + \log_2 n + \cdots
\end{align*}
The first term is the maximum number of bits to store the count $n_i$ of each of
the $k$ output values. The second term is the number of bits needed to specify
the particular observed sequence within the class of sequences that have counts
$n_1, n_2, \ldots, n_k$. The third term is the number $b$ of bits to specify the
number of bits in $n$ itself. Finally, the ellipsis indicates that we have to
specify the number of bits to specify $b$ ($\log_2 \log_2 n$) and so on, until
there is less than one bit.

We can make sense of this and so develop a helpful comparison to the original
storage estimate of $n \log_2 k$ bits, if we apply Stirling's approximation: $n!
\approx \sqrt{2 \pi n} \left(n/e\right)^n$. For a sufficiently long measurement
series, a little algebra gives:
\begin{align*}
\log_2 M &\approx -n \sum_{i=1}^k \frac{n_i}{n} \log_2 \frac{n_i}{n} \\
  &= n H[n_1/n, n_2/n, \ldots, n_k/n] ~.
\end{align*}
bits for $n$ observations. Here, the function $H[P]$ is Shannon's \emph{entropy}
of the distribution $P = (n_1/n, n_2/n, \ldots, n_k/n)$. As a shorthand, when
discussing the information in a random variable $X$ that is distributed
according to $P$, we also write $H[X]$. Thus, to the extent that $H[X] \le
\log_2 k$, as the series length $n$ grows the observer can effectively compress
the original series of observations and so use less storage than
$n \log_2 k$.

\begin{figure}
  \centering
  
  \ifgenerate
    \input{img/stdviewofbit.tikz}
  \else
    \includegraphics{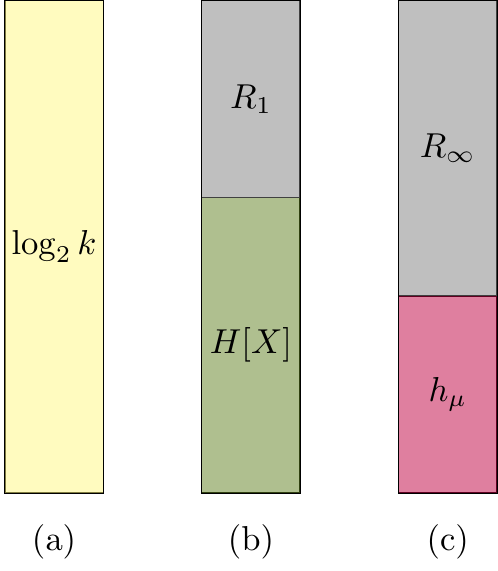}
  \fi

  \caption{Dissecting information in a single measurement $X$ being one of $k$
    values. }
\label{fig:StdViewOfBit}
\end{figure}

The relationship between the raw measurement ($\log_2 k$) and the average-case
view ($H[X]$), that we just laid out explicitly, is illustrated in the contrast
between Figs. \ref{fig:StdViewOfBit}(a) and \ref{fig:StdViewOfBit}(b). The
difference $R_1 = \log_2 k - H[X]$ is the amount of redundant information in the
raw measurements. As such, the magnitude of $R_1$ indicates how much they can be
compressed.

Information storage can be reduced further, since using $H[X]$ as the amount of
information in a measurement implicitly assumed the instrument's outputs were
statistically independent. And this, as it turns out, leads to $H[X]$ being an
overestimate as to the amount of information in $X$. For general information
sources, there are correlations and restrictions between successive measurements
that violate this independence assumption and, helpfully, we can use these to
further compress \emph{sequences} of measurements---$X_1, X_2, \ldots, X_\ell$.
Concretely, information theory tells us that the irreducible information per
observation is given by the Shannon \emph{entropy rate}:
\begin{equation}
\hmu = \lim_{\ell \to \infty} \frac{H(\ell)}{\ell} ~,
\end{equation}
where $H(\ell) = - \sum_{\{x^\ell\}} \Prob(x^\ell) \log_2 \Prob(x^\ell)$ is the
\emph{block entropy}---the Shannon entropy of the length-$\ell$ word
distribution $\Prob(x^\ell)$.

The improved view of the information in a measurement is given in Fig.
\ref{fig:StdViewOfBit}(c). Specifically, since $\hmu \leq H[X]$, we can compress
even more; indeed, by an amount $R_{\infty} = \log_2 k - \hmu$.

These comments are no more than a review of basic information theory
\cite{Cover2006} that used a little algebra. They do, however, set the stage for
a parallel, but more detailed, analysis of the information in an observation. In
focusing on a single measurement, the following complements recent, more
sophisticated analyses of information sources that focused on a process's hidden
states \cite[and references therein]{Crutchfield2010}. In the sense that the
latter is a state-centric informational analysis of a process, the following
takes the complementary measurement-centric view.

\begin{figure}
  \centering
  
  \ifgenerate
    \input{img/disection.tikz}
  \else
    \includegraphics{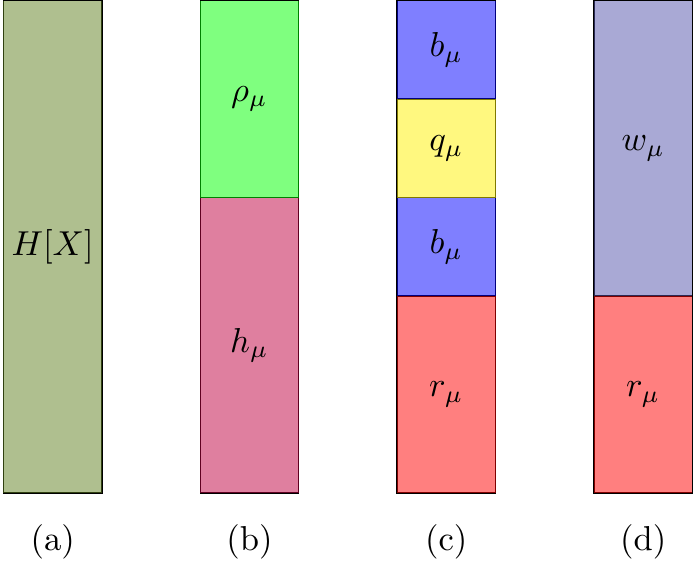}
  \fi

  \caption{Systematic dissection of $H[X]$. }
\label{fig:disection}
\end{figure}

Partly as preview and partly to orient ourselves on the path to be followed, we
illustrate the main results in a pictorial fashion similar to that just given;
see Fig. \ref{fig:disection} which further dissects the information in $X$.

As a first cut, the information $H[X]$ provided by each observation
(Fig.~\ref{fig:disection}(a))
can be broken into two pieces: one part is information $\rhomu$ that could be
anticipated from prior observations and the other $\hmu$---the random
component---is that which could not be anticipated.
(See Fig.~\ref{fig:disection}(b).)
Each of these pieces can be
further decomposed into two parts. The random component $\hmu$ breaks into two
kinds of randomness: a part $\bmu$ relevant for predicting the future, while the
remaining part $\rmu$ is ephemeral, existing only for the moment.

The redundant portion $\rhomu$ of $H[X]$ in turn splits into two pieces. The
first part---also $\bmu$ when the process is stationary---is shared between the
past and the current observation, but its relevance stops there. The second
piece $\qmu$ is anticipated by the past, is present currently, and also plays a
role in future behavior. Notably, this informational piece can be negative.
(See Fig.~\ref{fig:disection}(c).)

We can further combine all elements of $H[X]$ that participate in
structure---whether it be past, future, or both---into a single element $\wmu$.
This decomposition of $H[X]$ provides a very different decomposition than $\hmu$
and $\rhomu$. It partitions $H[X]$ into a piece $\wmu$ that is structural and a
piece $\rmu$ that, as mentioned above, is ephemeral.
(See Fig.~\ref{fig:disection}(d).)

With the basic informational components contained in a single measurement laid
out, we now derive them from first principles. The next step is to address
information in collections of random variables, helpful in a broad array of
problems. We then specialize to time series; viz., one-dimensional chains of
random variables.

\section{Information Measures}
\label{sec:information-measures}

Shannon's information theory~\cite{Cover2006} is a widely used mathematical
framework with many advantages in the study of complex, nonlinear systems. Most
importantly, it provides a unified quantitative way to analyze systems with
broadly dissimilar physical substrates. It further makes no assumptions as to
the types of correlation between variables, picking up multi-way nonlinear
interactions just as easily as simple pairwise linear correlations.

The workhorse of information theory is the Shannon \emph{entropy} of a random
variable, just introduced. The entropy measures what would commonly be
considered the amount of information learned, on average, from observing a
sample from that random variable. The entropy $H[X]$ of a random variable $X$ taking on
values $x \in \MeasAlphabet = \{1, \ldots, k\}$ with distribution $\Prob(X=x)$
has the following functional form:
\begin{align}
  H[X] = -\sum_{x \in \MeasAlphabet} \Prob(x) \log_2 \Prob(x) ~.
\label{eq:entropy}
\end{align}
The entropy is defined in the same manner over joint random variables---say, $X$
and $Y$---where the above distribution is replaced by the joint probability
$\Prob(X,Y)$.

When considering more than a single random variable, it is quite reasonable to
ask how much uncertainty remains in one variable given knowledge of the other.
The average entropy in one variable $X$ given the outcome of another variable
$Y$ is the \emph{conditional entropy}:
\begin{align}
  \label{eq:cond_ent}
  H[X|Y] = H[X,Y] - H[Y] ~.
\end{align}
That is, it is the entropy of the joint random variable $(X,Y)$ with the
marginal entropy $H[Y]$ of $Y$ subtracted from it.

The fundamental measure of correlation between random variables is the
\emph{mutual information}. As stated before, it can be adapted to measure all
kinds of interaction between two variables. It can be written in several forms,
including:
\begin{align}
  \label{eq:mi}
  I[X;Y] =& H[X] + H[Y] - H[X,Y] \\
         =& H[X,Y] - H[X|Y] - H[Y|X] ~.
\label{eq:MI}
\end{align}
Two variables are generally considered \emph{independent} if their mutual
information is zero.

Like the entropy, the mutual information can also be conditioned on another
variable, say $Z$, resulting in the \emph{conditional mutual information}. Its
definition is a straightforward modification of Eq.~(\ref{eq:mi}):
\begin{align}
  \label{eq:cmi}
  I[X;Y|Z] = H[X|Z] + H[Y|Z] - H[X,Y|Z] ~.
\end{align}

For example, consider two random variables $X$ and $Y$ that take the values $0$
or $1$ independently and uniformly, and a third $Z = X~\textbf{XOR}~Y$, the
exclusive-or of the two. There is a total of two bits of information among the
three variables: $H[X, Y, Z] = 2$ bits. Furthermore, the variables $X$ and $Y$
share
a single bit of information with $Z$, their parity. Thus, $I[X, Y; Z] = 1$ bit.
Interestingly, although $X$ and $Y$ are independent, $I[X;Y] = 0$, they are not
conditionally independent: $I[X;Y|Z] = 1$.

\section{Multivariate Information Measures}
\label{sec:multvariate-inform-measures}

We now turn to a difficult problem: How does one quantify interactions among an
arbitrary set of variables? As just noted, the mutual information provides a
very general, widely applicable method of measuring dependence between
\emph{two}, possibly composite, random variables. The challenge comes in the
fact that there exist several distinct methods for measuring dependence between
more than two random variables.

Consider a finite set $\MeasAlphabet$ and random variables $X_i$ taking on
values $x_i \in \MeasAlphabet$ for all $i \in \mathbb{Z}$. The vector of $N$
random variables $X_{0:N} = \{X_0, X_1, \ldots, X_{N-1}\}$ takes on values in
$\MeasAlphabet^N$. A straightforward generalization of Eq.~\eqref{eq:entropy}
yields the \emph{joint entropy}:
\begin{align}
  H[X_{0:N}] = -\sum_{\{x_{0:N}\}} \Prob(x_{0:N}) \log_2 \Prob(x_{0:N}) ~,
\label{eq:e}
\end{align}
which measures the total amount of information contained in the joint
distribution. From here onward, we suppress notating the set $\{ x_{0:N} \}$ of
realizations over which the sums are taken.

In generalizing the mutual information to arbitrary sets of variables, we make
use of power sets. We let $\Omega_N = \{0,1,\ldots,N-1\}$ denote the universal
set over the variable indices and define $P(N) =
\mathcal{P}\bigl(\Omega_N\bigr)$ as the power set over $\Omega_N$. Then, for any
set $A \in P(N)$, its complement is denoted $\bar{A} = \Omega_N \setminus A$.
Finally, we use a shorthand to refer to the set of random variables
corresponding to index set $A$:
\begin{align}
  X_A \equiv \{X_i : i \in A\} ~.
\end{align}

There are at least three extensions of the two-variable mutual information, each
based on a different interpretation of what its original definition intended.
The first is the \emph{multivariate mutual information} or
\emph{co-information}~\cite{Bell2003}: $I[X_0; X_1; \ldots; X_{N-1}]$. Denoted
$I[X_{0:N}]$, it is the amount of mutual information to which \emph{all}
variables contribute:
\begin{align}
  I[X_{0:N}]
    & = -\sum \Prob(x_{0:N})
        \log_2 \left( \quad\prod_{\mathclap{A \in P(N)}} \Prob(x_A)^{-1^{|v|}} \right)
        \nonumber \\
    \label{eq:mmi1}
    & = -\sum_{\mathclap{A \in P(N)}} (-1)^{|A|} H[X_A] \\
    \label{eq:mmi2}
    & = H[X_{0:N}] - \sum_{\mathclap{\underset{ 0 < |A| < N }{A \in P(N)}}}
        I[X_{\vphantom{\bar{A}}A} | X_{\bar{A}}] ~,
\end{align}
where, e.g., $I[X_{\{1,3,4\}} | X_{\{0,2\}}] = I[X_1 ; X_3 ; X_4 | X_0 , X_2]$. It can
be verified that Eq.~(\ref{eq:mmi1}) is a generalization of Eq.~(\ref{eq:mi}),
adding and subtracting all possible entropies according to the number of random
variables they include. The co-information has several interesting properties.
First, it can be negative, though a consistent interpretation of what this means
is still lacking in the literature. Second, this measure vanishes if \emph{any
  two} variables in the set are completely independent. (That is, they are
independent and also conditionally independent with respect to all subsets of
the other variables.) This is true regardless of interdependencies among the
other variables.

In the second interpretation, the mutual information is seen as the relative
entropy between a joint distribution and the product of its marginals.
Specifically, the starting point is:
\begin{align}
I[X;Y] = \sum \Prob(x,y) \log_2 \frac{\Prob(x,y)}{\Prob(x)\Prob(y)} ~,
\end{align}
which is simply a rewriting of Eq.~(\ref{eq:mi}). When generalized from this
form, we obtain the \emph{total correlation}~\cite{Watanabe1960}:
\begin{align}
  T[X_{0:N}]
  & = \sum \Prob(x_{0:N})
	\log_2 \left( \frac{\Prob(x_{0:N})}{\Prob(x_0)\ldots\Prob(x_N)} \right)
	\nonumber \\
  & = \sum_{\mathclap{\underset{|A| = 1}{A \in P(N)}}}
      H[X_{\vphantom{\bar{A}}A}] - H[X_{0:N}] ~.
\label{eq:tc}
\end{align}
The total correlation is sometimes referred to as the ``multi-information'',
though we refrain from using this ambiguous term. It differs from the prior
measure in many fundamental ways. To begin with, it is nonnegative. It also
differs in that if $X_0$ is independent of the others, then $T[X_{0:N}] =
T[X_{1:N}]$. Finally, it captures only the difference between individual
variables and the entire set. The role of two-way and higher interactions is
ignored as it leaves out the relative entropies between the entire set and
more-than-two-variable marginals. Indeed, this is a common problem. The total
correlation and the next measure miss or, at best, conflate $(n > 2)$-way
interactions.

The last extension stems from the view that mutual information is the joint
entropy minus all (single-variable) unshared information---that is, we start
from Eq.~(\ref{eq:MI}). When interpreted this way, the generalization is called
the \emph{binding information}~\cite{Abdallah2010}:
\begin{align}
  B[X_{0:N}] = H[X_{0:N}] -
  \sum_{\mathclap{\underset{|A| = 1}{A \in P(N)}}}
  H[X_{\vphantom{\bar{A}}A} | X_{\bar{A}}] ~.
\label{eq:bi}
\end{align}
Like the total correlation, the binding information is nonnegative and
independent random variables do not change its value. Note that $B[X_{0:N}]$ is
a first approximation to the multivariate information of Eq.~(\ref{eq:mmi1})
when the sets $A$ are restricted to singleton sets.

We next define three additional multivariate information measures that have not
been studied previously, but appear following a similar strategy. First, we have
the amount of information in individual variables that is not shared in any way.
This is the \emph{residual entropy}:
\begin{align}
  R[X_{0:N}] & = H[X_{0:N}] - B[X_{0:N}]      \nonumber \\
             & = \sum_{\mathclap{\underset{|A| = 1}{A \in P(N)}}}
                 H[X_{\vphantom{\bar{A}}A} | X_{\bar{A}}] ~.
\label{eq:re}
\end{align}
In a sense, it is an anti-mutual information: It measures the total amount of
randomness localized to an individual variable and so not correlated to that in
its peers.

Second, we can sum the total correlation and the binding information. Then we
have the \emph{local exogenous information}:
\begin{align}
  W[X_{0:N}] & = B[X_{0:N}] + T[X_{0:N}] \\
     & = \sum_{\mathclap{\underset{|A| = 1}{A \in P(N)}}}
         \left( H[X_{\vphantom{\bar{A}}A}] -
         H[X_{\vphantom{\bar{A}}A} | X_{\bar{A}}] \right) \\
     & = \sum_{\mathclap{\underset{|A| = 1}{A \in P(N)}}}
         I[X_{\vphantom{\bar{A}}A} ; X_{\bar{A}}]
 	~.
\label{eq:lei}
\end{align}
It is the amount of information in each variable that comes from its peers. It
is a ``very mutual'' information, one that discounts for the randomness produced
locally---that randomness inherent in each variable individually.

$W[X_{0:N}]$ is close to the binding information, except that it uses the sum of
marginals not the joint entropy. As such, it seems to more consistently capture
the role of single variables within a set than $B[X_{0:N}]$, which compares the
set's joint entropy to individual residual uncertainties.

Third and finally, there is a measure which, for lack of a better name, we call
the \emph{enigmatic information}:
\begin{align}
  Q[X_{0:N}] & = T[X_{0:N}] - B[X_{0:N}]
  ~.
\label{eq:ei}
\end{align}
Like the multivariate mutual information---which it equals when $N=3$---it can
be negative. Its operational meaning will become clear on further discussion.

\section{Time Series}
\label{sec:time-series}

We now adapt the general multivariate measures to analyze discrete-valued,
discrete-time series generated by a stationary process. That is, rather than
analyzing sets of random variables, we specialize to a one-dimensional chain of
them. In this setting, the measures are most appropriately applied to
successively longer blocks of consecutive observations. This allows us to study
the asymptotic block-length behavior of each, mimicking the approach of Ref.
\cite{Crutchfield2003,Crutchfield2010}. For the class of processes known as
\emph{finitary} (defined shortly), each of these measures tend to a linear
asymptote characterized by a subextensive component and an extensive component
controlled by an asymptotic growth rate.

\begin{figure}
  \centering
  
  \ifgenerate
    \input{img/timeseries.tikz}
  \else
    \includegraphics{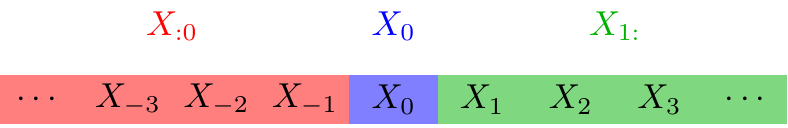}
  \fi

\caption{A process's time series: Time indices less than zero refer to the
  past $\Past$; index $0$ to the present $X_0$; and times after $0$ to the
  future $\MeasSymbols{1}{}$.
  }
\label{fig:timeseries}
\end{figure}

Let's first state more precisely and introduce the notation for the class of
processes that are the object of study. We consider a bi-infinite chain $\ldots
X_{-1} X_0 X_1 \ldots$ of random variables. Each $X_t, t \in \mathbb{Z}$, takes
on a finite set of values $x_t \in \MeasAlphabet$. We denote contiguous subsets
of the time series with $X_{A:B}$ where the left index is inclusive and the right
is exclusive. By leaving one of the indices off the subset is partially infinite
in that direction. We divide this bi-infinite chain into three segments. First
we single out the \emph{present} $X_0$. All the symbols prior to the present
are the \emph{past} $\Past$. The symbols following the present are the
\emph{future} $\MeasSymbols{1}{}$. Figure~\ref{fig:timeseries}
illustrates the setting.

Our focus is on the $\ell$-blocks $X_{t:t+\ell} = X_t X_{t+1} \cdots
X_{t+\ell-1}$. The associated \emph{process} is specified by the set of
length-$\ell$ word distributions: $\{ \Prob(X_{t:t+\ell}): t \in \mathbb{Z},
\ell \in \mathbb{N} \}$. We consider only \emph{stationary} processes for which
$\Prob(X_{t:t+\ell}) = \Prob(X_{0:\ell})$. And so, we drop the absolute-time
index $t$. More precisely, the word probabilities derive from an underlying
time-shift invariant, ergodic measure $\mu$ on the space of bi-infinite
sequences.

In the following, an information measure $\mathcal{F}$ applied to to the
process's length-$\ell$ words is denoted $\mathcal{F}[X_{0:\ell}]$ or, as a
shorthand, $\mathcal{F}(\ell)$.

\subsection{Block Entropy versus Total Correlation}
\label{sec:relat-betw-entr}

We begin with the long-studied block entropy information measure $H(\ell)$
~\cite{Crutchfield1983, Eriksson1987}. (For a review and background to the
following see Ref.~\cite{Crutchfield2003}.) The block entropy curve defines two
primary features. First, its growth rate limits to the \emph{entropy rate}
$\hmu$. Second, its subextensive component is the \emph{excess entropy} $\EE$:
\begin{align}
	\EE = I[\Past;\Future] ~,
\label{eq:ee}
\end{align}
which expresses the totality of information shared between the past and future.

The entropy rate and excess entropy, and the way in which they are approached
with increasing block length, are commonly used quantifiers for complexity in
many fields. They are complementary in the sense that, for finitary processes,
the block entropy for sufficiently long blocks takes the form:
\begin{align}
  \label{eq:be}
  H(\ell) \sim \EE + \ell \hmu ~.
\end{align}
Recall that $H(0) = 0$ and that $H(\ell)$ is monotone increasing and concave
down. The finitary processes, mentioned above, are those with finite $\EE$.

\newcommand{\BTC}{T}

Next, we turn to a less well studied measure for time series---the \emph{block
  total correlation} $T(\ell)$. Adapting Eq.~(\ref{eq:tc}) to a stationary
process gives its definition:
\begin{align}
  \BTC(\ell) = \ell H[X_0] - H(\ell) ~.
\label{eq:btc}
\end{align}
Note that $\BTC(0) = 0$ and $\BTC(1) = 0$. Effectively, it compares a process's
block entropy to the case of independent, identically distributed random
variables. In many ways, the block total correlation is the reverse side of an
information-theoretic coin for which the block entropy is the obverse. For
finitary processes, its growth rate limits to a constant $\rhomu$ and its
subextensive part is a constant that turns out to be $-\EE$:
\begin{align}
  \BTC(\ell) \sim -\EE + \ell \rhomu ~.
\label{eq:btc_scaling}
\end{align}
That is, $\rhomu = \lim_{\ell \to \infty} \BTC(\ell)/\ell$. Finally,
$\BTC(\ell)$ is monotone increasing, but concave up. All of this is derived
directly from Eqs.~(\ref{eq:be}) and (\ref{eq:btc}), by using well known
properties of the block entropy.

\begin{figure}
  \centering
  \includegraphics{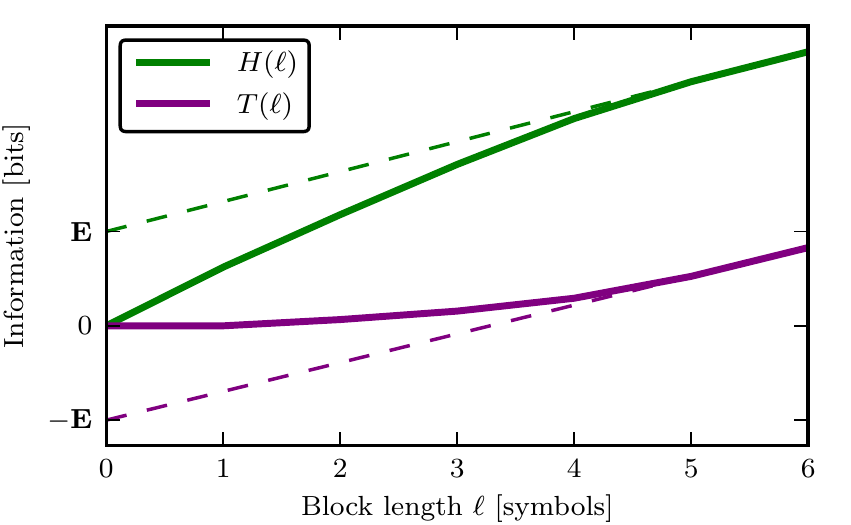}
  \caption{Block entropy $H(\ell)$ and block total correlation $\BTC(\ell)$
    illustrating their behaviors for the NRPS Process. }
\label{fig:be-btc}
\end{figure}

The block entropy and block total correlation are plotted in
Fig.~\ref{fig:be-btc}. Both measures are $0$ at $\ell = 0$ and from there
approach their asymptotic behavior, denoted by the dashed lines. Though their
asymptotic slopes appear to be the same, they in fact differ. Numerical
data for the asymptotic values can be found in Tables \ref{tab:decomp} and
\ref{tab:ee} under the heading NRPS (defined later).

There is a persistent confusion in the neuroscience, complex systems, and
information theory literatures concerning the relationship between block entropy
and block total correlation. This can be alleviated by explicitly demonstrating
a partial symmetry between the two in the time series setting and by
highlighting a weakness of the total correlation.

We begin by showing how, for stationary processes, the block entropy and the
block total correlation contain much the same information. From
Eqs.~(\ref{eq:e}) and (\ref{eq:tc}) we immediately see that:
\begin{align}
  \label{eq:sum-be-btc}
  H(\ell) + \BTC(\ell) = \ell H(1) ~.
\end{align}
Furthermore, by substituting Eqs.~(\ref{eq:be}) and (\ref{eq:btc_scaling}) in
Eq.~(\ref{eq:sum-be-btc}) we note that the righthand side has no subextensive
component. This gives further proof that the subextensive components of
Eqs.~(\ref{eq:be}) and (\ref{eq:btc_scaling}) must be equal and opposite, as
claimed. Moreover, by equating individual $\ell$-terms we find:
\begin{align}
  \label{eq:h-plus-rho}
  \hmu + \rhomu = H(1) ~.
\end{align}
And, this is the decomposition given in Fig.~\ref{fig:disection}(b): the
lefthand side provides two pieces comprising the single-observation entropy
$H(1)$.

Continuing, either information measure can be used to obtain the excess entropy.
In addition, since the block entropy provides $\hmu$ as well as intrinsically
containing $H(1)$, $\rhomu$ can be directly obtained from the block entropy
function by taking $H(1) - \hmu$, yielding $\rhomu$. The same is not true,
however, for the total correlation. Though $\rhomu$ can be computed, one cannot
obtain $\hmu$ from $\BTC(\ell)$ alone---$H(1)$ is required, but not available
from $\BTC(\ell)$, since it is subtracted out.

There are further parallels between the two quantities that can be drawn. First,
following Ref. \cite{Crutchfield2003}, we define discrete derivatives of the
block measures at length $\ell$:
\begin{align}
  h_\ell    & = H(\ell) - H(\ell - 1) \\
  \rho_\ell & = \BTC(\ell) - \BTC(\ell - 1) ~.
  \label{eq:hmul-rhomul}
\end{align}
These approach $\hmu$ and $\rhomu$, respectively. From them we can determine the
subextensive components by discrete integration, while subtracting out the
asymptotic behavior. We find that:
\begin{align}
  \EE &= \phantom{-}\sum_{\ell = 1}^\infty \left( h_\ell - \hmu \right)
  \label{eq:sums1}
\intertext{and also that}
  \EE &= -\sum_{\ell = 1}^\infty \left( \rho_\ell - \rhomu \right) ~.
  \label{eq:sums2}
\end{align}
Second, these sums are equal term by term.

The first sum, however, indirectly brings us back to Eq.~(\ref{eq:h-plus-rho}).
Since $h_1 = H(1)$, we have:
\begin{align}
  \EE = \rhomu + \sum_{\ell = 2}^\infty \left( h_\ell - \hmu \right) ~.
\label{eq:rho-in-E}
\end{align}

Finally, it has been said that the total correlation (``multi-information'')
is the first term in $\EE$~\cite{Ay2006, Erb2004}. This has perhaps given the
impression that the total correlation is only useful as a crude approximation.
Equation~(\ref{eq:rho-in-E}) shows that it is actually the total correlation
\emph{rate} $\rhomu$ that is $\EE$'s first term. As we just showed, the total
correlation is more useful than being a first term in an expansion. Its utility
is ultimately limited, though, since its properties are redundant with that of
the block entropy which, in addition, gives the process's entropy rate $\hmu$.

\subsection{A Finer Decomposition}
\label{sec:finer-decomposition}

We now show how, in the time series setting, the binding information, local
exogenous information, enigmatic information, and residual entropy constitute a
refinement of the single-measurement decomposition provided by the block entropy and the total
correlation~\cite{Abdallah2010, Abdallah2010a}. To begin, their block
equivalents are, respectively:
\begin{align}
  B(\ell) &= H(\ell) - R(\ell) \\
  Q(\ell) &= \BTC(\ell) - B(\ell) \\
  W(\ell) &= B(\ell) + \BTC(\ell) ~,
\label{eq:B-W-Q}
\end{align}
where $R(\ell)$ does not have an analogously simple form. Their asymptotic
behaviors are, respectively:
\begin{align}
  \label{eq:B-R-Q}
  R(\ell) &\sim \EER + \ell \rmu \\
  B(\ell) &\sim \EEB + \ell \bmu \\
  Q(\ell) &\sim \EEQ + \ell \qmu \\
  W(\ell) &\sim \EEW + \ell \wmu ~.
\end{align}
Their associated rates break the prior two components ($\hmu$ and $\rhomu$) into
finer pieces. Substituting their definitions into Eqs.~(\ref{eq:e}) and
(\ref{eq:btc}) we have:
\begin{align}
  \label{eq:be-btc-breakdown}
  H(\ell)    &= B(\ell) + R(\ell) \\
	         &= (\EEB + \EER) + \ell (\bmu + \rmu)
			   \label{eq:be-breakdown-rates} \\
  \BTC(\ell) &= B(\ell) + Q(\ell) \\
  	         &= (\EEB + \EEQ) + \ell (\bmu + \qmu) ~.
			   \label{eq:btc-breakdown-rates}
\end{align}
The rates in Eqs.~(\ref{eq:be-breakdown-rates}) and
(\ref{eq:btc-breakdown-rates}) corresponding to $\hmu$ and $\rhomu$,
respectively, give the decomposition laid out in Fig.~\ref{fig:disection}(c)
above. Two of these components ($\bmu$ and $\rmu$) were defined in Ref.
\cite{Abdallah2010} and the third ($\qmu$) is a direct extension. We defer
interpreting them to Sec.~\ref{sec:process-information-diagram} which provides
greater understanding by appealing to the semantics afforded by the process
information diagram developed there.

The local exogenous information, rather than refining the decomposition provided
by the block entropy and the total correlation, provides a different
decomposition:
\begin{align}
  \label{eq:blei-breakdown}
  W(\ell) =& B(\ell) + \BTC(\ell) \\
          =& (\EEB - \EE) + \ell (\bmu + \rhomu) ~.
\end{align}
So, $\wmu = \hmu + \rhomu$, as mentioned in Fig.~\ref{fig:disection}(d).

Similar to Eq.~(\ref{eq:sum-be-btc}), we can take the local exogenous
information together with the residual entropy and find:
\begin{align}
  \label{eq:sum-r-w}
  R(\ell) + W(\ell) = \ell H(1) ~.
\end{align}
This implies that $\EER = -\EEW$ and that $\rmu$ and $\wmu$ are yet another
partitioning of $H[X]$, as shown earlier in Fig.~\ref{fig:disection}(d).

\begin{figure}
  \centering
  \includegraphics{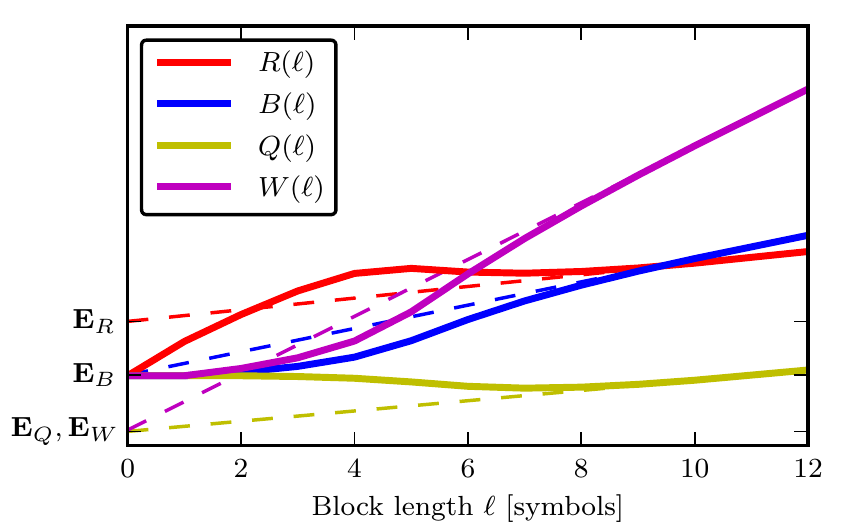}
  \caption{Block equivalents of the residual entropy $R(\ell)$, binding
    information $B(\ell)$, enigmatic information $Q(\ell)$, and local exogenous
    information $W(\ell)$ for a generic process (same as previous figure). }
\label{fig:bre-bbi-bei}
\end{figure}

Figure~\ref{fig:bre-bbi-bei} illustrates these four block measures for a generic
process. Each of the four measures reaches asymptotic linear behavior at a
length of $\ell = 9$ symbols. Once there, we see that they each possess a slope
that we just showed to be a decomposition of the slopes from the measures in
Fig.~\ref{fig:be-btc}. Furthermore, each has a subextensive component that is
found as the $y$-intercept of the linear asymptote. These subextensive parts
provide a decomposition of the excess entropy, discussed further below in
Sec.~\ref{sec:decompositions-ee}.

\subsection{Multivariate Mutual Information}
\label{sec:mult-mutu-inform}

Lastly, we come to the block equivalent of the multivariate mutual information
$I[X_{0:N}]$:
\begin{align}
  I(\ell) = H(\ell)
    - \sum_{\mathclap{\underset{0 < |A| < \ell}{A \in P(\ell)}}}
	I[X_{\vphantom{\bar{A}}A} | X_{\bar{A}}] ~.
\label{eq:bmmi}
\end{align}
Superficially, it scales similarly to the other measures:
\begin{align}
  \label{eq:bmmi_scaling}
  I(\ell) \sim \I + \ell \imu ~,
\end{align}
with an asymptotic growth rate $\imu$ and a constant subextensive component
$\I$. Yet, it has differing implications regarding what it captures in the
process. This is drawn out by the following propositions, whose proofs appear
elsewhere.

The first concerns the subextensive part of $I(\ell)$.
\begin{Prop}
For all finite-state processes:
  \begin{align}
    \hmu > 0 \quad \Rightarrow \quad \lim_{\ell \to \infty} I(\ell) = 0 ~.
    \label{eq:mmi-zero}
  \end{align}
\label{prop:mmi-zero}
\end{Prop}
The intuition behind this is fairly straightforward. For $I(\ell)$ to be
nonzero, no two observations can be independent. Finite-state processes with
positive $\hmu$ are stochastic, however. So, observations become (conditionally)
decoupled exponentially fast. Thus, for arbitrarily long blocks, the first and
the last observations tend toward independence exponentially and so $I(\ell)$
limits to $0$.

The second proposition regards the growth rate $\imu$.
\begin{Prop}
For all finite-state processes:
  \begin{align}
    \label{eq:imu-e-zero}
    \imu = 0 ~.
  \end{align}
\label{prop:imu}
\end{Prop}
The intuition behind this follows from the first proposition. If $\hmu > 0$,
then it is clear that since $I(\ell)$ tends toward $0$, then the slope must also
tend toward $0$. What remains are those processes that are finite state but for
which $\hmu = 0$. These are the periodic processes. For them, $\imu$ also
vanishes since, although $I(\ell)$ may be nonzero, there is a finite amount of
information contained in a bi-infinite periodic sequence. Once all this
information has been accounted for at a particular block length, then for all
blocks larger than this there is no additional information to gain. And so,
$\imu$ decays to $0$.

The final result concerns the subextensive component $\I$.
\begin{Prop}
For all finite-state processes with $\hmu~>~0$:
  \begin{align}
    \label{eq:I-e-zero}
    \I = 0 ~.
  \end{align}
\label{prop:eei}
\end{Prop}
This follows directly from the previous two propositions.

Thus, the block multivariate mutual information is qualitatively different from
the other block measures. It appears to be most interesting for infinitary
processes with infinite excess entropy.

\begin{figure}
  \centering
  \includegraphics{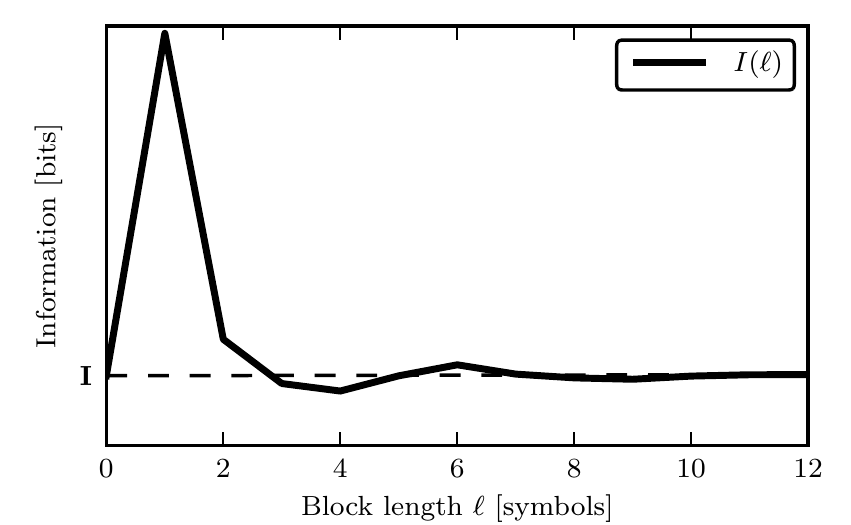}
  \caption{Block multivariate mutual information $I(\ell)$ for the same
  	example process as before.
	}
\label{fig:bmmi}
\end{figure}

Figure~\ref{fig:bmmi} demonstrates the general behavior of $I(\ell)$,
illustrating the three propositions. The dashed line highlights the asymptotic
behavior of $I(\ell)$: both $\I$ and $\imu$ vanish. We further see that
$I(\ell)$ is not restricted to positive values. It oscillates about $0$ until
length $\ell = 11$ where it finally vanishes.

\section{Information Diagrams}
\label{sec:information-diagram}

Information diagrams~\cite{Yeung1991} provide a graphical and intuitive way to
interpret the information-theoretic relationships among variables. In
construction and concept, they are very similar to Venn diagrams. The key
difference is that the measure used is a Shannon entropy rather than a set size.
Additionally, an overlap is not set intersection but rather a mutual
information. The irreducible intersections are, in fact, elementary \emph{atoms}
of a sigma-algebra over the random-variable event space. An atom's size reflects
the magnitude of one or another \emph{Shannon information measure}---marginal,
joint, or conditional entropy or mutual information.

\subsection{Four-Variable Information Diagrams}
\label{sec:four-variable-information-diagram}

\begin{figure}
  \centering
  \subfigure[Joint entropy, Eq.~(\ref{eq:e})]{
    
  \ifgenerate
    \input{img/subfig-id1.tikz}
  \else
    \includegraphics{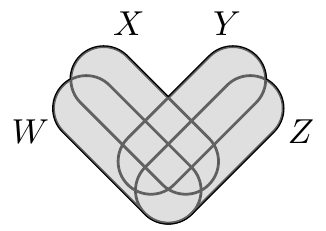}
  \fi

    \label{subfig:id1}
  } \\ 
  \subfigure[Multivariate mutual information, Eq.~(\ref{eq:mmi1})]{
    
  \ifgenerate
    \input{img/subfig-id2.tikz}
  \else
    \includegraphics{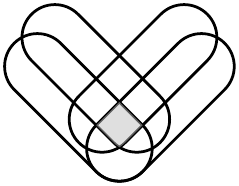}
  \fi

    \label{subfig:id2}
  }
  \subfigure[Total correlation, Eq.~(\ref{eq:tc})]{
    
  \ifgenerate
    \input{img/subfig-id3.tikz}
  \else
    \includegraphics{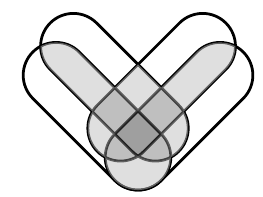}
  \fi

    \label{subfig:id3}
  }
  \subfigure[Binding information, Eq.~(\ref{eq:bi})]{
    
  \ifgenerate
    \input{img/subfig-id4.tikz}
  \else
    \includegraphics{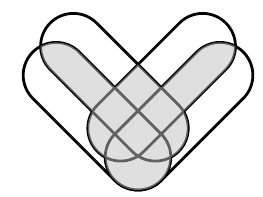}
  \fi

    \label{subfig:id4}
  }
  \subfigure[Residual entropy, Eq.~(\ref{eq:re})]{
    
  \ifgenerate
    \input{img/subfig-id5.tikz}
  \else
    \includegraphics{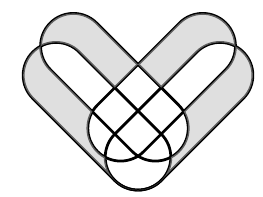}
  \fi

    \label{subfig:id5}
  }
  \subfigure[Local exogenous information, Eq.~(\ref{eq:lei})]{
    
  \ifgenerate
    \input{img/subfig-id6.tikz}
  \else
    \includegraphics{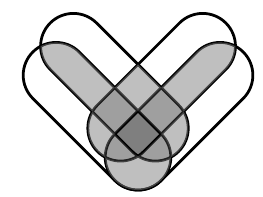}
  \fi

    \label{subfig:id6}
  }
  \subfigure[Enigmatic information, Eq.~(\ref{eq:ei})]{
    
  \ifgenerate
    \input{img/subfig-id7.tikz}
  \else
    \includegraphics{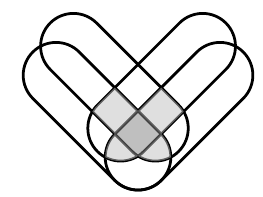}
  \fi

    \label{subfig:id7}
  }
  \caption{Four-variable information diagrams for the multivariate
    information measures of Sec.~\protect\ref{sec:multvariate-inform-measures}.
    Darker shades of gray denote heavier weighting in the corresponding
    informational sum. For example, the atoms to which all four variables
    contribute are added thrice to the total correlation and so the central
    atom's weight $I[W;X;Y;Z] = 3$.}
\label{fig:fourvariablemeasures}
\end{figure}

Using information diagrams we can deepen our understanding of the multivariate
informations defined in Sec.~\ref{sec:multvariate-inform-measures}.
Fig.~\ref{fig:fourvariablemeasures} illustrates them for four random
variables---$X$, $Y$, $Z$, $W$. There, an atom's shade of gray denotes how much
weight it carries in the overall value of its measure. Consider for example the
total correlation I-diagram in Fig.~\ref{fig:fourvariablemeasures}(c). From the
definition of the total correlation, Eq.~(\ref{eq:tc}), we see that each
variable provides one count to each of its atoms and then a count is removed
from each atom. Thus, the atom associated with four-way intersection $W \cap X
\cap Y \cap Z$ contained in each of the four variables carries a total weight
$I[W;X;Y;Z] = 4 - 1 = 3$. Those atoms contained in three variables carry a
weight of $2$, those shared among only two variables a weight of $1$, and
information solely contained in one variable is not counted at all.

Utilizing the I-diagrams in Fig.~\ref{fig:fourvariablemeasures}, we can easily
visualize and intuit how these various information measures relate to each other
and the distributions they represent. In Fig.~\ref{subfig:id1}, we find the
joint entropy. Since it represents all information contained in the distribution
with no bias to any sort of interaction, we see that it counts each and every
atom once. The residual entropy, Fig.~\ref{subfig:id5}, is equally easy to
interpret: it counts each atom which is not shared by two or more variables.

The distinctions in the menagerie of measures attempting to capture interactions
among $N$ variables can also be easily seen. The multivariate mutual
information, Fig.~\ref{subfig:id2}, stands out in that it is isolated to a
single atom, that contained in all variables. This makes it clear why the
independence of any two of the variables leads to a zero value for this measure.
The total correlation, Fig.~\ref{subfig:id3}, contains all atoms contained in at
least two variables and gives higher weight to those contained in more
variables. The local exogenous information, Fig.~\ref{subfig:id6}, is similar.
It counts the same atoms as the total correlation does, but it gives them higher
weight. Lastly, the binding information, Fig.~\ref{subfig:id4}, also counts the
same atoms, but only weights each of them once regardless of how many variables
they participate in.

The lone enigmatic information, Fig.~\ref{subfig:id7}, counts only those
variables that participate in at least three variables and, similar to the total
correlation, it counts those that participate in more variables more heavily.

\subsection{Process Information Diagrams}
\label{sec:process-information-diagram}

\begin{figure}
  \centering
  
  \ifgenerate
    \input{img/IDiagramAnatomy.tikz}
  \else
    \includegraphics{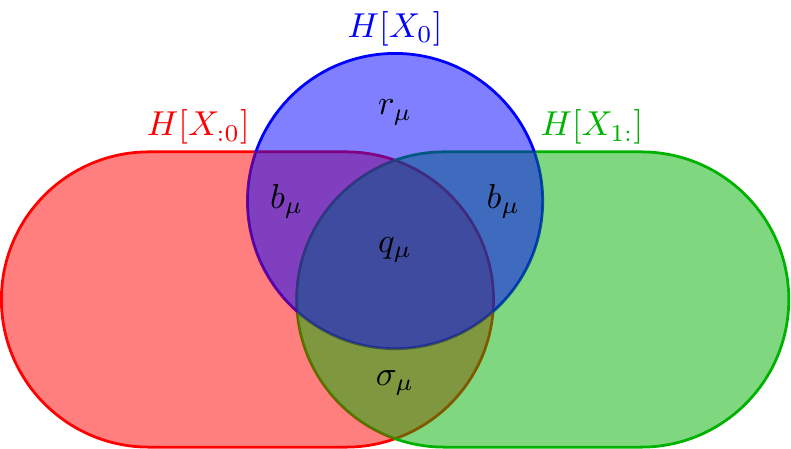}
  \fi

  \caption{I-diagram anatomy of $H[\MeasSymbol_0]$ in the full context of
    time: The past $\Past$ partitions $H[\MeasSymbol_0]$ into two pieces: $\hmu$
    and $\rhomu$. The future $\Future$ then partitions those further into
    $\rmu$, two $\bmu$s, and $\qmu$. This leaves a component $\sigmamu$, shared
    by the past and the future, that is not in the present $\MeasSymbol_0$.}
\label{fig:anatomy}
\end{figure}

Following Ref.~\cite{Crutchfield2009} we adapt the multivariate I-diagrams just
laid out to tracking information in finitary stationary processes. In
particular, we develop process I-diagrams to explain the information in a single
observation, as described before in Fig.~\ref{fig:disection}. The resulting
process I-diagram is displayed in Fig.~\ref{fig:anatomy}. As we will see,
exploring the diagram gives a greater, semantic understanding of the
relationships among the process variables and, as we will emphasize, of the
internal structure of the process itself.

\begin{figure}
  \centering
  \subfigure[]{
    
  \ifgenerate
    \input{img/subfig-decomp1.tikz}
  \else
    \includegraphics{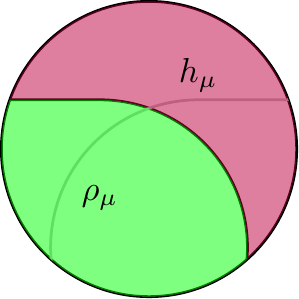}
  \fi

    \label{subfig:decomp1}
  }
  \subfigure[]{
    
  \ifgenerate
    \input{img/subfig-decomp2.tikz}
  \else
    \includegraphics{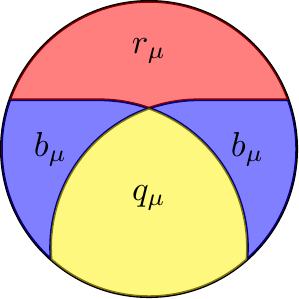}
  \fi

    \label{subfig:decomp2}
  }
  \subfigure[]{
    
  \ifgenerate
    \input{img/subfig-decomp3.tikz}
  \else
    \includegraphics{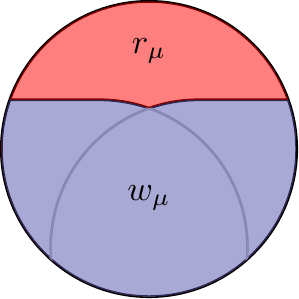}
  \fi

    \label{subfig:decomp3}
  }
  \caption{The three decompositions of $H[X]$ from Fig.~\ref{fig:disection}. The
    dissecting lines are identical to those in Fig.~\ref{fig:anatomy}.}
\end{figure}

For all measures, except the multivariate mutual information, the extensive rate
corresponds to one or more atoms in the decomposition of $H[X_0]$. To begin, we
allow $H[X_0]$ to be split in two by the past. This exposes two pieces: $\hmu$,
the part exterior to the past, and $\rhomu$, the part interior. This
partitioning has been well studied in information theory due to how it naturally
arises as one observes a sequence. This decomposition is displayed in
Fig.~\ref{subfig:decomp1}.

Taking a step back and including the future in the diagram, we obtain a more
detailed understanding of how information is transmitted in a process. The past
and the future together divide $H[X_0]$ into four parts; see
Fig.~\ref{subfig:decomp2}. We will discuss each part shortly. First, however, we
draw out a different decomposition---that into $\rmu$ and $\wmu$ as seen in
Fig.~\ref{subfig:decomp3}. From this diagram it is easy to see the semantic
meaning behind the decomposition: $\rmu$ being divorced from any temporal
structure, while $\wmu$ is steeped in it.

We finally turn to the partitioning shown in Fig.~\ref{subfig:decomp2}. The
process I-diagram makes it rather transparent in which sense $\rmu$ is an amount
of \emph{ephemeral} information: its atom lies outside both the past and future
sets and so it exists only in the present moment, having no repercussions for
the future and being no consequence of the past. It is the amount of information
in the present observation neither communicated to the future nor from the past.
Ref.~\cite{Abdallah2010} referred to this as the residual entropy \emph{rate},
as it is the amount of uncertainty that remains in the present even after
accounting for every other variable in the time series.

Ref.~\cite{Abdallah2010} also proposed to use $\bmu$ as a measure of structural
complexity~\cite{Abdallah2010}, and we tend to agree. The argument for this is
intuitive: $\bmu$ is an amount of information that is present now, is not
explained by the past, but has repercussions in the future. That is, it is the
portion of the entropy rate $\hmu$ that has consequences. In some contexts one
may prefer to employ the ratio $\bmu/\hmu$ when $\bmu$ is interpreted an
indicator of complex behavior since, for a fixed $\bmu$, larger $\hmu$ values
imply less temporal structure in the time series.

Due to stationarity, the mutual information $I[X_0;\MeasSymbols{1}{}|\Past]$
between the present $X_0$ and the future $\MeasSymbols{1}{}$ conditioned on the
past $\Past$ is the same as the mutual information
$I[X_0;\Past|\MeasSymbols{1}{}]$ between $X_0$ and the past $\Past$ conditioned
on the future $\MeasSymbols{1}{}$. Moreover, both are $\bmu$. This lends a
symmetry to the process I-diagram that does not exist for nonstationary
processes. Thus, $\bmu$ atoms in Fig.~\ref{fig:anatomy} are the same size.

There are two atoms remaining in the process I-diagram that have not been
discussed in literature. Both merit attention. The first is $\qmu$---the
information shared by the past, the present, and the future. Notably, its value
can be negative and we discuss this further below in
Sec.~\ref{sec:negativity-q_mu}. The other piece, denoted $\sigma_\mu$, is a
component of information shared between the past and the future that
\emph{does not exist in the present} observation. This piece is vital evidence
that attempting to understand a process without using a model for its generating
mechanism is ultimately incomplete. We discuss this point further in
Sec.~\ref{sec:cons-sigm} below.

\subsubsection{Negativity of $\qmu$}
\label{sec:negativity-q_mu}

\newcommand{\partinfo}[1]{\Pi_{\{#1\}}}
\newcommand{\redundancy}{\partinfo{\Past\}\{\MeasSymbols{1}{}}}
\newcommand{\synergy}{\partinfo{\Past, \MeasSymbols{1}{}}}
\newcommand{\piiota}{\partinfo{\Past}, \partinfo{\MeasSymbols{1}{}}}

The sign of $\qmu$ holds valuable information. To see what this is we apply the
partial information decomposition \cite{Williams2010} to further analyze $\wmu =
I[\MeasSymbol_0;\Past,\MeasSymbols{1}{}]$---that portion of the present shared
with the past and future. By decomposing $\wmu$ into four pieces---three of
which are unique---we gain greater insight into the value of $\qmu$ and also
draw out potential asymmetries between the past and the future.

The partial information lattice provides us with a method to isolate (i) the
contributions $\redundancy$ to $\wmu$ that both the past and the future provide
redundantly, (ii) parts $\partinfo{\Past}$ and $\partinfo{\MeasSymbols{1}{}}$
that are uniquely provided by the past and the future, respectively, and (iii) a
part $\synergy$ that is synergistically provided by both the past and the
future. Note that, due to stationarity, $\partinfo{\Past}
= \partinfo{\MeasSymbols{1}{}}$. We refer to this as the \emph{uniquity} and
denote it $\iota$.

Using Ref.~\cite{Williams2010} we see that $\qmu$ is equal to the
\emph{redundancy} minus the \emph{synergy} of the past and the future, when
determining the present. Thus, if $\qmu > 0$, the past and future predominantly
contribute information to the present. When $\qmu < 0$, however, considering the
past and the future separately in determining the present misses essential
correlations. The latter can be teased out if the past and future are considered
together.

The process I-diagram (Fig.~\ref{fig:anatomy}) showed that the mutual
information between the present and either the past or the future is $\rhomu$.
One might suspect from this that the past and the future provide the same
information to the present, but this would be incorrect. Though they provide the
same \emph{quantity} of information to the present, \emph{what} that information
conveys can differ. This is evidence of a process's structural irreversibility;
cf. Refs.~\cite{Crutchfield2009,Ellison2009}. In this light, the
redundancy $\redundancy$ between the past and future when considering the
present is $\rhomu - \iota$. Furthermore, the synergy $\synergy$ provided by the
past and the future is equal to $\bmu - \iota$.

Taking this all together, we find what we already knew: that $\qmu = \rhomu -
\bmu$, The journey to this conclusion, however, provided us with deeper insight
into what negative $\qmu$ means and into the structure of $\wmu$ and the process as a whole.

\begin{figure}
  \centering
  
  \ifgenerate
    \input{img/partial_info.tikz}
  \else
    \includegraphics{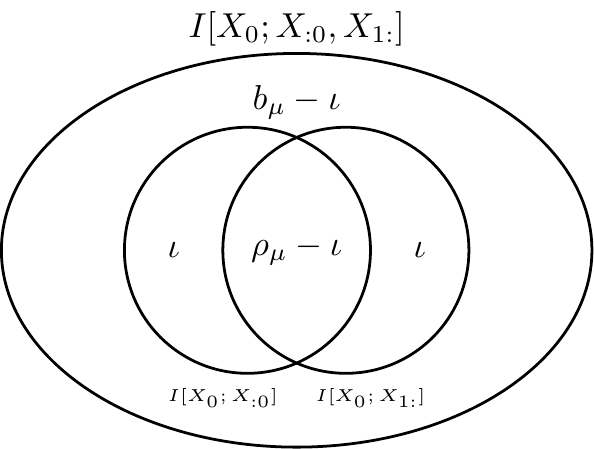}
  \fi

  \caption{Partial information decomposition of $\wmu = I[ X_0; X_{:0}, X_{1:}
    ]$. The multivariate mutual information $\qmu$ is given by the redundancy
    $\redundancy$ minus the synergy $\synergy$. $\wmu = \rhomu + \bmu$ is the
	sum of all atoms in this diagram.
	}
\label{fig:partial_info}
\end{figure}

\subsubsection{Consequence of $\sigmamu$: Why we model}
\label{sec:cons-sigm}

Notably, the final piece of the process I-diagram is not part of $H[X_0]$---not
a component of the information in a single observation. This is $\sigmamu$,
which represents information that is transmitted from the past to the future,
but does not go through the currently observed symbol $\MeasSymbol_0$. This is
readily understood and leads to an important conclusion.

If one believes that the process under study is generated according to the laws
of physics, then the process's internal physical configuration must store all
the information from the past that is relevant for generating future behavior.
Only when the observed process is order-$1$ Markov is it sufficient to keep
track of just the current observable. For the plethora of processes that are not
order-$1$ or that are non-Markovian altogether, we are faced with the fact that
information relevant for future behavior must be stored somehow. And, this fact
is reflected in the existence of $\sigma_\mu$. When $\sigmamu > 0$, a complete
description of the process requires accounting for this internal configurational
or, simply, \emph{state information}. This is why we build models and cannot
rely on only collecting observation sequences.

The amount of information shared between $\Past$ and $\MeasSymbols{1}{}$, but
ignoring $\MeasSymbol_0$, was previously discussed in Ref.~\cite{Ball2010}. We
now see that the meaning of this information quantity---there denoted
$\mathcal{I}_1$---is easily gleaned from its components: $\mathcal{I}_1 = \qmu +
\sigmamu$.

Furthermore, in Refs. \cite{Abdallah2010}, \cite{Abdallah2010a}, and
\cite{Ball2010}, efficient computation of $\bmu$ and $\mathcal{I}_1$ were not
provided and the brute force estimates are inaccurate and very compute
intensive. Fortunately, by a direct extension of the methods developed in
Ref.~\cite{Ellison2009} on bidirectional machines, we can easily compute both
$\rmu = H[X_0 | \CausalState_0^+, \CausalState_1^-]$ and $\mathcal{I}_1 =
I[\CausalState_0^+, \CausalState_1^-]$. This is done by constructing joint
probabilities of forward-time and reverse-time causal
states---$\{\FutureCausalState\}$ and $\{\PastCausalState\}$,
respectively---at different time indices employing the dynamic
of the bidirectional machine. This gives closed-form, exact methods of
calculating these two measures, provided one constructs the process's forward
and reverse \eMs. $\bmu$ follows directly in this case since it is the
difference of $\hmu$ and $\rmu$; the former is also directly calculated from the \eM.

\subsubsection{Decompositions of $\EE$}
\label{sec:decompositions-ee}

Using the process I-diagram and the tools provided above, three unique
decompositions of the excess entropy, Eq. ~(\ref{eq:ee}), can be given. Each
provides a different interpretation of how information is transmitted from the
past to the future.

The first is provided by
Eqs.~(\ref{eq:be-btc-breakdown})-(\ref{eq:btc-breakdown-rates}). The
subextensive parts of the block entropy and total correlation there determine
the excess entropy decomposition. We have:
\begin{align}
  \label{eq:1}
  \EE =&              \EEB + \EER  \\
      =&            - \EEB - \EEQ  \\
      =&  \frac{1}{2}(\EER - \EEQ) \\
      =& -\frac{1}{2}(\EEW + \EEQ) ~.
\end{align}
We leave the meaning behind these decompositions as an open problem, but do note
that they are distinct from those discussed next.

The second and third decompositions both derive directly from the process
I-diagram of Fig.~\ref{fig:anatomy}. Without further work, one can easily see
that the excess entropy breaks into three pieces, all previously discussed:
\begin{align}
  \label{eq:2}
  \EE = \bmu + \qmu + \sigmamu ~.
\end{align}

And, finally, one can perform the partial information decomposition on the
mutual information $I[\Past; X_0, \MeasSymbols{1}{}]$. The result gives an
improved understanding of (i) how much information is uniquely shared with the
either the immediate or the more distant future and (ii) how much is redundantly
or synergistically with both.

The decompositions provided by the atoms of the process I-diagram and those
provided by the subextensive rates of block-information curves are conceptually
quite different. It has been shown \cite{Feldman2003} that the subextensive part
of the block entropy and the mutual information between the past and the future,
though equal for one dimensional processes, differ in two dimensions. We believe
the semantic differences shown here are evidence that the degeneracy of
alternate $\EE$-decompositions breaks in higher dimensions.

\section{Examples}
\label{sec:examples}

\begin{figure}
  \centering
  \subfigure[Even Process]{
    
  \ifgenerate
    \input{img/even.tikz}
  \else
    \includegraphics{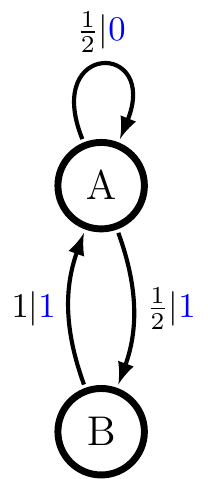}
  \fi

  }
  \subfigure[Golden Mean Process]{
    
  \ifgenerate
    \input{img/goldenmean.tikz}
  \else
    \includegraphics{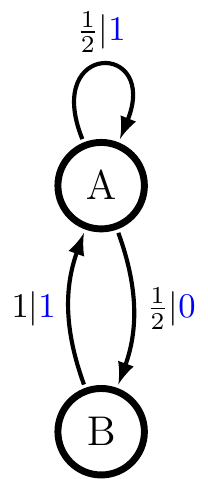}
  \fi

  }
  \subfigure[Noisy Random Phase-Slip Process]{
    
  \ifgenerate
    \input{img/nrps.tikz}
  \else
    \includegraphics{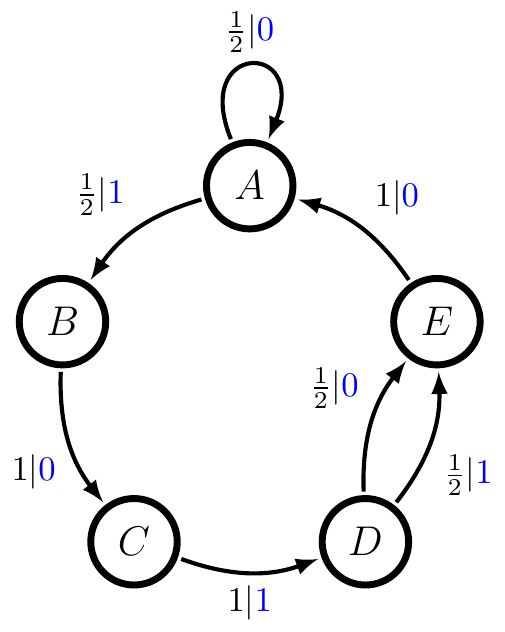}
  \fi

  }
  \caption{\EM\ presentations for the three example processes.}
  \label{fig:ems}
\end{figure}

We now make the preceding concrete by calculating these quantities for three
different processes, selected to illustrate a variety of informational
properties. Figure \ref{fig:ems} gives each process via it's \eM\
\cite{Shalizi2001}: the Even Process, the Golden Mean Process, and the Noisy
Random Phase-Slip (NRPS) Process. A process's \eM\ consists of its \emph{causal
  states}---a partitioning of infinite pasts into sets that give rise to the
same predictions about future behavior. The state transitions are labeled
$\Edge{p}{s}$ where $\Symbol{s}$ is the observed symbol and $p$ is the
conditional probability of observing that symbol given the state the process is
in. The \eM\ representation for a process is its minimal unifilar presentation.

Table~\ref{tab:decomp} begins by showing the single-observation entropy $H[1]$
followed by $\hmu$ and $\rhomu$. Note that the Even and the Golden Mean
Processes cannot be differentiated using these measures alone. The table then
follows with the finer decomposition. We now see that the processes can be
differentiated. We can understand fairly easily that the Even Process, being
infinite-order Markovian, and consisting of blocks of $1$s of even length
separated by one or more $0$s, exhibits more structure than the Golden Mean
Process. (This is rather intuitive if one recalls that the Golden Mean Process
has only a single restriction: it cannot generate sequences with consecutive
$0$s.) We see that, for the Even Process, $\rmu$ is $0$. This can be understood
by considering a bi-infinite sample from the Even Process with a single gap in
it. The structure of this process is such that we can always and immediately
identify what that missing symbol must be.

These two processes are further differentiated by $\qmu$, where it is negative
for the Even Process and positive for the Golden Mean Process. On the one hand,
this implies that there is a larger amount of synergy than redundancy in the
Even Process. Indeed, it is often the case, when appealing only to the past or
the future, that one cannot determine the value of $X_0$, but when taken
together the possibilities are limited to a single symbol. On the other hand,
since $\qmu$ is positive for the Golden Mean Process we can determine that its
behavior is dominated by redundant contributions. That $\wmu$ is larger for the
Even Process than the Golden Mean Process is consonant with the impression that
the former is, overall, more structured.

The next value in the table is $\sigmamu$, the amount of state
information not contained in the current observable. This vanishes for the
Golden Mean Process, as it is order-$1$ Markovian. The Even Process, however,
has a significant amount of information stored that is not observable in the
present.

Last in the table is a partial information decomposition of $I[X_0; \Past,
\MeasSymbols{1}{}]$. $\qmu$ is given by $\redundancy - \synergy$. Of note here
is that the NRPS process's nonzero uniquity $\iota = 0.02437$. For the Even and
Golden Mean Processes it vanishes. That is, in the NRPS Process information is
uniquely communicated to the present from the past and an equivalent in
magnitude, but different, information is communicated to the future. Thus, the
NRPS Process illustrates a subtle asymmetry in statistical structure.

\begin{table}
  \centering
  \begin{tabular}{l|r|r|r|}
    \cline{2-4}
                                        & Even     & Golden Mean & NRPS     \\
    \hline
    \multicolumn{1}{|l|}{$H[1]$}        &  0.91830 &  0.91830    &  0.97987 \\
    \hline
    \multicolumn{1}{|l|}{$\hmu$}        &  0.66667 &  0.66667    &  0.50000 \\
    \multicolumn{1}{|l|}{$\rhomu$}      &  0.25163 &  0.25163    &  0.47987 \\
    \hline
    \multicolumn{1}{|l|}{$\rmu$}        &  0.00000 &  0.45915    &  0.16667 \\
    \multicolumn{1}{|l|}{$\bmu$}        &  0.66667 &  0.20752    &  0.33333 \\
    \multicolumn{1}{|l|}{$\qmu$}        & -0.41504 &  0.04411    &  0.14654 \\
    \multicolumn{1}{|l|}{$\wmu$}        &  0.91830 &  0.45915    &  0.81320 \\
    \hline
    \multicolumn{1}{|l|}{$\sigmamu$}    &  0.66667 &  0.00000    &  1.09407 \\
    \hline
    \multicolumn{1}{|l|}{$\redundancy$} &  0.25163 &  0.25163    &  0.45550 \\
    \multicolumn{1}{|l|}{$\iota: \piiota$}     &  0.00000 &  0.00000    &  0.02437 \\
    \multicolumn{1}{|l|}{$\synergy$}    &  0.66667 &  0.20752    &  0.30896 \\
    \hline
  \end{tabular}
  \caption{Information measure analysis of three processes.}
  \label{tab:decomp}
\end{table}

Table~\ref{tab:ee} then provides an alternate breakdown of $\EE$ for each
prototype process. We use this here to only highlight how much the processes
differ in character from one another. The consequences of the first
decomposition of excess entropy---$\EE = \bmu + \qmu + \sigmamu$---follow
directly from the previous table's discussion. The second and third
decompositions into $\EER + \EEB$ and $-\EEB - \EEQ$ vary from one another
significantly. The Even Process has much larger values for these pieces than the
total $\EE$, whereas the NRPS process has two values nearly equal to $\EE$ and
one very small. The Golden Mean Process falls somewhere between these two.

The final excess entropy breakdown is provided by the partial information
decomposition of $I[\Past; X_0, \MeasSymbols{1}{}]$. Here, we again see
differing properties among the three processes. The Even Process consists only
of redundancy $\redundancy$ and synergy $\synergy$. The Golden Mean Process
contains no synergy, a small amount of redundancy, and most of its information
sharing is with the present uniquely. The NRPS Process possesses both synergy
and redundancy, but also a significant amount of information shared solely with
the more distant future.

\newcommand{\EEred}{\partinfo{X_0\}\{\MeasSymbols{1}{}}}
\newcommand{\EEpre}{\partinfo{X_0}}
\newcommand{\EEfut}{\partinfo{\MeasSymbols{1}{}}}
\newcommand{\EEsyn}{\partinfo{X_0, \MeasSymbols{1}{}}}

\begin{table}
  \centering
  \begin{tabular}{l|r|r|r|}
    \cline{2-4}
                                       & Even     & Golden Mean & NRPS     \\
    \hline
    \multicolumn{1}{|l|}{$\EE$}        &  0.91830 &  0.25163    &  1.57393 \\
    \hline
    \multicolumn{1}{|l|}{$\bmu$}       &  0.66667 &  0.20752    &  0.33333 \\
    \multicolumn{1}{|l|}{$\qmu$}       & -0.41504 &  0.04411    &  0.14654 \\
    \multicolumn{1}{|l|}{$\sigmamu$}   &  0.66667 &  0.00000    &  1.09407 \\
    \hline
    \multicolumn{1}{|l|}{$\EER$}       &  4.48470 &  0.41504    &  1.55445 \\
    \multicolumn{1}{|l|}{$\EEB$}       & -3.56640 & -0.16341    &  0.01948 \\
    \multicolumn{1}{|l|}{$\EEQ$}       &  2.64810 & -0.08822    & -1.59342 \\
    \multicolumn{1}{|l|}{$\EEW$}       & -4.48470 & -0.41504    & -1.55445 \\
    \hline
    \multicolumn{1}{|l|}{$\EEred$}     &  0.25163 &  0.04411    &  0.47987 \\
    \multicolumn{1}{|l|}{$\EEpre$}     &  0.00000 &  0.20752    &  0.00000 \\
    \multicolumn{1}{|l|}{$\EEfut$}     &  0.00000 &  0.00000    &  0.76073 \\
    \multicolumn{1}{|l|}{$\EEsyn$}     &  0.66667 &  0.00000    &  0.33333 \\
    \hline
  \end{tabular}
  \caption{Alternative decompositions of excess entropy $\EE$ for
    the three prototype processes.}
  \label{tab:ee}
\end{table}

\begin{figure}
  \centering
  \includegraphics{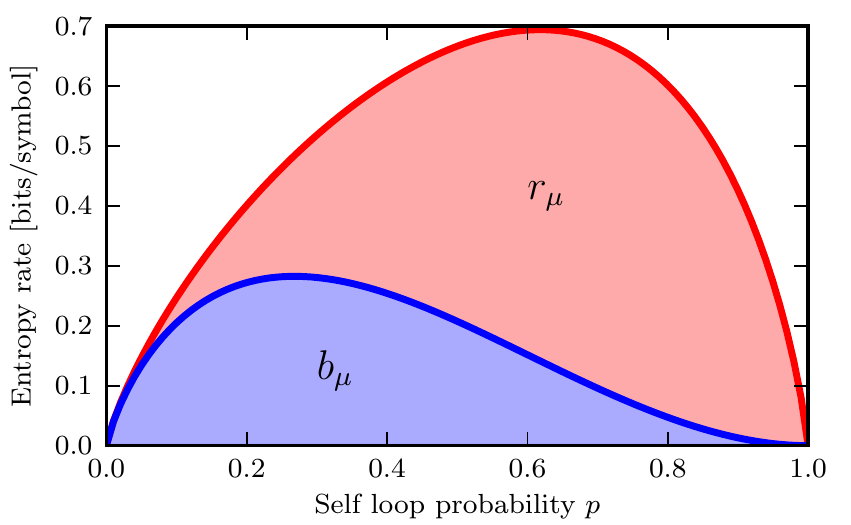}
  \caption{The breakdown of $\hmu$ for the Golden Mean Process. The self-loop
    probability was varied from $0$ to $1$, adjusting the other edge's
    probability accordingly. }
\label{fig:rmu-bmu}
\end{figure}

And, finally, Fig.~\ref{fig:rmu-bmu} plots how $\hmu$ partitions into $\rmu$ and
$\bmu$ for the Golden Mean family of processes. This family consists of all
processes with \eM\ structure given in Fig.~\ref{fig:ems}(b), but where the
outgoing transition probabilities from state \textbf{A} are parametrized. We can
easily see that for small self-loop transition probabilities, the majority of
$\hmu$ is consumed by $\bmu$. This should be intuitive since, when the self-loop
probability is small, the process is nearly periodic and $\rmu$ should be nearly
zero. On the other end of the spectrum, when the self-loop probability is large,
$\hmu$ is mostly consumed by $\rmu$. This is again intuitive since observations
from that process are dominated by $1$s and the occasional $0$---which provides
all the entropy for $\hmu$---has no effect on structure.

\section{Concluding Remarks}
\label{sec:conclusion}

We began by outlining a conceptual decomposition of a single observation in a
time series: a single observation contains a hierarchy of informational
components. We then made the decomposition concrete using a variety of
multivariate information measures. Adapting them to time series, we showed that
their asymptotic growth rates are identified with the hierarchical
decomposition. To unify the various competing views, we provided the
measurement-centric process I-diagram, demonstrating that it concisely reveals
the semantic meaning behind each component in the hierarchy.

Once the measurement-centric process I-diagram was available, we isolated two
components, analyzing in detail their meaning. We utilized the partial
information lattice~\cite{Williams2010} to refine our understanding of when the
past and the future redundantly and synergistically inform the present. This
allowed us to explain a subtle statistical asymmetry---the directionality in the
difference between $\rhomu$ and $\redundancy$.

The other atom we singled out in the process I-diagram was $\sigmamu$. It is the
most compelling evidence that analyzing a process from its measurements alone,
without constructing a state-based model, is ultimately limited.

Next, we discussed how the different methods and measures relate to one of the
most widely used complexity measures---the past-future mutual information or
excess entropy. In particular, we showed how they yield four distinct
decompositions and, in some cases, give useful interpretations of what these
decompositions mean operationally.

Then, we calculated all the measures for three different prototype processes,
each highlighting particular features of the information-theoretic
decompositions. We gave interpretations of negative mutual informations, as seen
in $\qmu$. The interpretations were consistent, understandable, and insightful.
There was nothing untoward about negative informations.

By adapting it to the time series setting, we highlighted a key weakness of the
total correlation (or multi-information). This undoubtedly explains the lack of
interest in using it in the time series setting, though the weakness still holds
when it is used to analyze any group of random variables. The weakness has led
to persistent over-interpretations of what it describes. It also may have
eclipsed the importance of its more complete analog, such as the block entropy,
in the settings of networked random variables.

In closing, we take a longer view. There is an exponential number of possible
atoms for $N$-way information measures. In addition, there is a similarly large
number possible partial information decompositions for $N$ variables. This
diversity presents the possibility of a large number of independent efforts
to define and uniquely motivate why one or the other information measure is
the best. Indeed, many of these yet-to-be-explored measures may be useful. In
this light, there is a bright future for developing information measures
adapted to a wide range of nonlinear, complex systems.
And, helpfully, a unifying framework appears to be emerging.

\section*{Acknowledgments}

We thank John Mahoney, Nick Travers, and J\"{o}rg Reichardt for many helpful
discussions and feedback. This work was partially supported by the Defense
Advanced Research Projects Agency (DARPA) Physical Intelligence Subcontract No.
9060-000709. The views, opinions, and findings contained in this article are
those of the authors and should not be interpreted as representing the official
views or policies, either expressed or implied, of the DARPA or the Department
of Defense.

\bibliography{anatomy}

\end{document}